\documentclass[runningheads,fleqn]{cl2emult}
\usepackage{makeidx}  % allows index generation
\usepackage{graphicx} % standard LaTeX graphics tool
\usepackage{subeqnar} % subnumbers individual equations
\usepackage{multicol} % used for the two-column index
\usepackage{cropmark} % cropmarks for pages without
\usepackage{phys}     % flushleft layout of math and captions
\makeindex            % used for the subject index
\begin{document}
\title*{Incipient Spanning Clusters \protect\newline
in Square and Cubic Percolation}
\toctitle{Incipient Spanning Clusters \protect\newline
in Square and Cubic Percolation}
% allows explicit linebreak for the table of content
%
%
\titlerunning{Incipient Spanning Clusters}
% allows abbreviation of title, if the full title is too long
% to fit in the running head
%
\author{Lev N. Shchur}
\authorrunning{Lev N. Shchur}
% if there are more than two authors,
% please abbreviate author list for running head
%
%
\institute{Landau Institute for Theoretical Physics,
142432 Chernogolovka, Russia}

\maketitle              % typesets the title of the contribution

\begin{abstract}

The analysis of extensive numerical data for the percolation probabilities
of incipient spanning clusters in two dimensional percolation at
criticality are presented. We developed an effective code for the
single-scan version of the Hoshen-Kopelman algorithm.  We measured the
probabilities on the square lattice forming samples of rectangular strips
with widths from 8 to 256 sites and lengths up to 3200 sites. At total of
more than $10^{15}$ random numbers are generated for the sampling
procedure. Our data confirm the proposed exact formulaes for the
probability exponents conjectured recently on the base of 2D conformal
field theory. Some preliminary results for 3D percolation are also
discussed.\footnote{Invited Talk at 12th Annual Workshop on Computer
Simulation Studies in Condensed Matter Physics, Athens, GA, March 8-12,
1999}

\end{abstract}

\section{Introduction}

 Percolation is a quite common phenomena in nature. Among known examples
there are epidemic diseases of garden trees, forest fires, the process of
espresso preparation, Ising spins at criticality, etc. \cite{Bunde}. Ising
spins are correlated with probability $p=1-\exp(-2J/k_BT)$ ($J$ is
coupling constant, $k_B$ - Boltzmann factor and $T$ - temperature) and
clusters of parallel spins percolate at the critical temperature $T_c$
\cite{FK}.
 
 Percolation is the simplest example of a critical phenomenon. Scaling
relations for infinite lattices were developed in the same way as for the
Ising model \cite{Stauf-Ar}. Mapping of the percolation model onto a $q=1$
state Potts model \cite{FK} gave a way to predict critical exponents for
percolation in two dimensions using techniques of Conformal Field Theory
(CFT) or of Coulomb gas representation.

 In the last few years, new insights into the properties of percolation
clusters on infinite lattices at the critical concentration have been
developed. First of all, Langlands and coauthors found numerically
\cite{Lang} that the probability of a cluster spanning the lattice
horizontally is an universal function of the aspect ratio only, and does
not depend on the lattice symmetry or on the type of percolation (see also
\cite{Ziff92}).
 
Second, Aizenman proved that there could be more than one incipient
spanning cluster in two-dimensional critical percolation \cite{Aiz1}. His
results were in contradiction with a wide-spread belief that the
percolation cluster is unique (see, for example \cite{Grim}).  Numerical
results for percolation in a strip demonstrated that there could be more
than one cluster spanning the short direction in a strip \cite{Hu}.  
Later Aizenman's proposal was supported numerically for the case of bond
percolation \cite{SK1} (see also \cite{Sen}).

Finally, Cardy conjectured an exact form for the probability that $k$
clusters span a large strip or open cylinder of aspect ratio $r$, in the
limit when $k$ is large \cite{Cardy98}. His result is in a complete
agreement with the Aizenman theorem.

The assumptions of conformal field theory are not rigorously established,
and it remains important to perform precise numerical tests of the theory.
I present here an extensive numerical verification of Cardy's proposal.
The numerical data were obtained in collaboration with Sergey Kosyakov
\cite{SK1,Topos,SK-fs}.

In Section~2 some known facts on percolation probabilities are shortly
reviewed. Spanning probability are discussed in Section~3. Most recent
findings on the multiplicity of incipient spanning clusters (ISC), are
discussed in Section~4, where our previously published results for the ISC
probabilities in bond percolation are compared with the proposed exact
result. The computational method is described in the Section~5. New
results for the finite size corrections of spanning probabilities in a
strip geometry with up to 819200 sites are presented in Sections~6 and 7
for open boundaries and cylinder geometry. Preliminary results for the
spanning probabilities in a simple cubic critical percolation with the
samples with up to 1900544 sites are presented in Section~8.  All averages
were done using $10^8$ strips in all geometries investigated.  Discussion
and acknowledgements finish the paper.

\section{Percolation}

\begin{figure*}
\centering
\includegraphics[width=3in,height=3in]{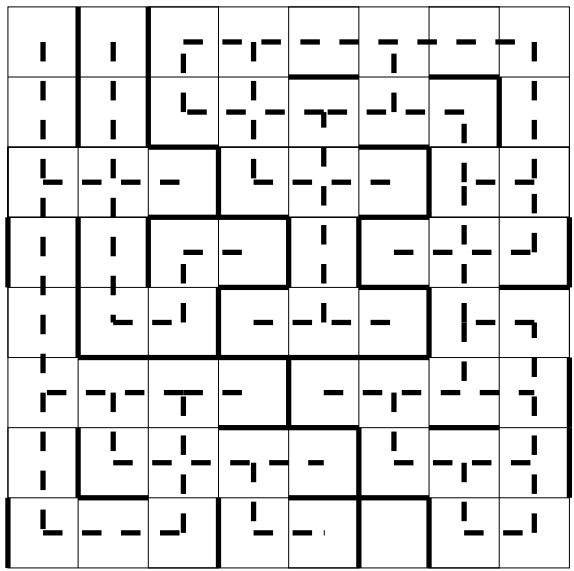}
\caption{Example of a prime lattice and its dual. Closed bonds of the
prime lattice are denoted by bold full lines, and those of the dual
lattice by bold dashed lines. Open prime bonds are plotted with thin full
lines. Open dual bonds are not shown.}
\label{dual}
\end{figure*}

The simplest model of percolation is bond percolation on the square
lattice.  Let us consider a two dimensional (2D) rectangle with an
infinite number of sites $L$ in both (vertical and horizontal) directions.
The probability for the bond to be occupied (or placed, or black) is equal
to $p$, $0\le p \le 1$. Consequently, the probability for the bond to be
open (or removed, or white) is equal to $q=1-p$. If all bonds are closed
(that is $p=1$) the left side is connected with the right one by closed
bonds. That is closed, paths from the left boundary to the right boundary
exist. Clearly, if all bonds are open ones ($p=0$) there is no way to go
by closed bonds through the rectangle. It appears that there is some
value of the probability $p_c$ dividing these two possibilities. The value
of critical probability $p_c$ is equal to $1/2$ for bond percolation in
two dimensions \cite{Essam-80,Kesten}.  This can be seen from the
self-duality of the 2D square lattice. 

The dual lattice is constructed placing open bonds across the closed bonds
on the prime lattice, and by drawing closed dual bonds across the open
prime bonds. An example configuration is plotted in Fig.~\ref{dual}. It is
clear from the construction that the number of closed prime bonds is equal
to the number of open dual bonds. If $p$ is the probability that a bond is
closed on the prime lattice, then $q=1-p$ is the probability that the
chosen bond be closed on the dual lattice.  For a given probability $p$ of
closed bonds, let us denote by $P_h(p)$ the probability that there is a
path of connected closed bonds connecting left and right sides of the
prime lattice. That is $P_h(p)$ is the probability of a cluster spanning
the lattice horizontally. By $P_v^*(q)$ we will denote probability that a
cluster of closed bonds spans the dual lattice vertically for a given
probability $q$ of closed dual bonds. Two relations follows

\begin{eqnarray} P_h(p)+P_v^*(q) & = & 1 \nonumber \\
P_v(p)+P_h^*(q) & = & 1. 
\label{Phv} 
\end{eqnarray}
   
The prime and dual lattices are identical in this construction. Therefore,
the percolation threshold should be at $p_c=q_c$, i.e. $p_c=1/2$. A strong
proof of that statement is the main subject of Kesten's book
\cite{Kesten}.

\begin{figure*}
\centering
\includegraphics[width=3in,height=3in]{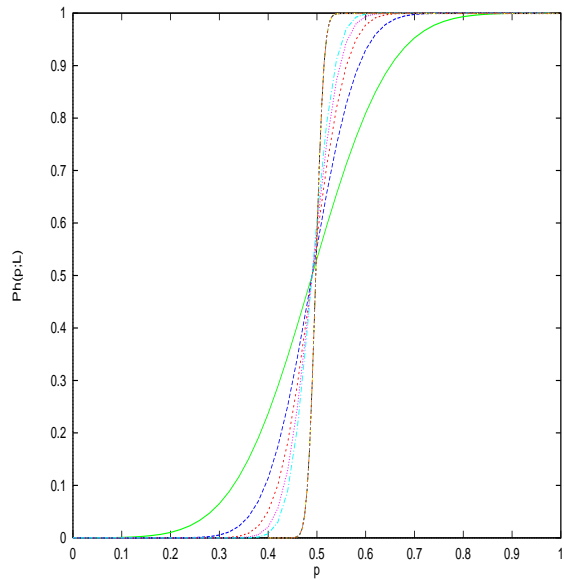}
\caption{Spanning probability $\pi_h(p,L)$ as function of bond occupation
probability $p$ for several values of lattice sizes $L=10,20,30,40,50$ and
$200$. The larger lattice size the larger derivative of spanning
probability.}
\label{step}
\end{figure*}

For the finite system with square size $L\times L$ the probability
$P_h(p;L)$ of a horizontal crossing is shown on Fig.~\ref{step} as a
function of $p$ for several $L$. The slope of this curve grows with the
lattice size $L$ as $L^{1/\nu}$, $\nu$ being the correlation length
exponent \cite{Stauf-Ar}. Is is clearly visible that $P_h(p;L)$ approaches
a step function as $L$ tends to infinity. Indeed, there are strong
mathematical results \cite{Kesten,AizB} supporting this picture.
 
What is the value of the spanning probability exactly at the percolation
threshold $p_c$? It was believed for a long time that this value was equal
to the value of the percolation threshold \cite{Stauf-Ar}. Indeed, this is
true for the bond percolation on the square lattice. From relations
(\ref{Phv})  for the bond percolation on the infinite square lattice it is
clear that
 
\begin{equation}
P_h(p_c)=P_v^*(q_c)=P_v(p_c)=P_h^*(q_c)=1/2.  
\label{pi12}
\end{equation} 

At the same time, the real space renormalization analysis \cite{RKS} are
based on the relation $P_h(p)=p$ for the site percolation. In the last
case $p_c\approx 0.592746$, which is fairly close to 0.5 and reasonable
results for the critical exponents were computed. However it was
recognized later that the value of spanning probability $P_h(p_c)$ does
not depend on the type of percolation as the percolation threshold does.
Moreover it turns out that this value is invariant and depends only on the
aspect ratio of the rectangle \cite{Lang,Cardy92,Lang-Bull}.
 
\section{Incipient spanning probability}

All results described in this section as well as in the following sections
are obtained for critical percolation. The probability $p$ for the bond
(or site) to be closed is equal to the corresponding critical value $p_c$
throughout the remaining text. The value of $p_c$ is not universal and
depends on the dimensionality, lattice type (square, honeycomb, etc.) and
percolation type (site, bond, etc.).

The spanning probability appears to be universal and invariant under shape
transformations of the rectangle, also depends on the boundary conditions
\cite{Lang,Cardy92,Lang-Bull} (see also the mini review by Stauffer
\cite{mini-review}).

Let us consider a strip with vertical size $L_v=L$ and horizontal size
$L_h=rL$, where $r=L_h/L_v$ is the aspect ratio. Incipient spanning
clusters connect two segments on the boundary of a macroscopically large
strip. Langlands et al. \cite{Lang,Lang-Bull} found numerically evidence
that the probability that a such cluster exists is invariant under
conformal transformation.  Cardy conjectured an exact formula for this
probability \cite{Cardy92} in terms of hypergeometric functions

$$ 
P_h(\eta)=\frac{3\Gamma(\frac23)}{\Gamma(\frac13)^2} \eta^\frac13\;
_2F_1(\frac13, \frac23, \frac43 ;\eta),
\label{exact1}
$$ 
where $\Gamma$ and $_2F_1$ are the gamma function and hypergeometric
function respectively and $\eta=((1-k)/(1+k))^2$, where $k$ defines the
aspect ratio $r$ of the rectangle $r=K(1-k^2)/2K(k^2)$ as the ratio of two
complete elliptic integrals $K(u)$. The modulus $k$ is associated with the
positions $(-1/k,-1,1,1/k)$ on the real axes mapped under a
Schwartz-Christoffel transformation to the vertices $x_i$ of rectangle.  
Thus, the probability $P_h$ is just the probability that there are closed
paths (Incipient Spanning Clusters) which connect the interval $(x1,x2)$
with the opposite one $(x3,x4)$. This conjecture of Cardy was confirmed
numerically by Langlands, et al. \cite{Lang-Bull}.

A more practical form of spanning probability was developed by Ziff as
series expansion in powers of $\exp(-2\pi\; r)$ \cite{Ziff-C}
\begin{eqnarray}
P_h(r)  =& 2^\frac43 c\; (e^{-\pi r/3}-\frac47 e^{-7\pi r/3}
+\frac{2}{13} e^{-13\pi r/3}...), & r\ge 1 \\
               =& 1-2^\frac43 c\; (e^{-\pi /3r}-\frac47 e^{-7\pi /3r}
+\frac{2}{13} e^{-13\pi /3r}...), & r\le 1 
\label{z14ab}
\end{eqnarray}
where $c\equiv\Gamma(\frac23)/\Gamma(\frac13)^2\approx 0.566046680...$.
Note, the corrections $\exp(-2\pi\; r)$ to the leading behaviour
$\exp(-2\pi\; r/3)$ 
decay very fast with increasing $r$. 

The probability that at least one cluster spans the square with open
boundaries in both directions is equal to $1/2$. This value could be
obtained from (\ref{exact1}) or (\ref{z14ab}) at $r=1$.

In the case of cylinder the spanning probability along the cylinder is not
known exactly. Simulations by Hovi and Aharony \cite{HoviAr} give the
value $P_h(1)=0.6366(8)$ for the aspect ratio $r=1$.

In a forthcoming paper \cite{SK-fs} we compute the spanning
probabilities $P_h(r)$ for the aspect ratio $r$ in the interval $0.25\le
r\le 25$ for cylindrical geometry as well in the interval $0.25\le r \le
12.5$ in open geometry with the step $\Delta r=0.25$. The quality of
these results will be discussed in the following sections.

\section{Coexistence of Incipient Spanning Clusters in 2D}

In the following sections we will also compute the probabilities that at
least two, and even more clusters simultaneously span the lattice from
left to right.

 It was a common belief until recently that percolation clusters are
unique on the 2D lattice. Aizenman proposed quite recently in his lecture
at the Statistical Physics 19 Conference in China \cite{Aiz-19} that the
number of Incipient Spanning Clusters (ISC) in 2D can be larger than one.
Later he proved \cite{Aiz1} that the probability $P(k,r;L)$ that at least
$k$ ISCs span horizontally (that is along $L_h$) the strip
$(L_h,L_v)=(rL,r)$ with width $L_v$ and horizontal length $L_h$ bounded

\begin{equation}
A e^{-\alpha k^2r} \le P(k,r;L) \le e^{-\alpha' k^2r}, 
\label{Aiz-prop}
\end{equation}
where $\alpha$ and $\alpha'$ are different
positive constants expected to merge for an infinite lattice \cite{Aiz1}.
Indications of the existence of simultaneous clusters in two-dimensional
critical percolation in the limit of infinite lattices were found in
computer simulations by Sen\cite{Sen} for site percolation on square
lattices with helical boundary conditions and in a short strips by Hu and
Lin \cite{Hu} using their  Monte Carlo histogram method.

We  checked numerically the number of spanning clusters in the critical
bond percolation model on two-dimensional square lattices \cite{SK1}. We
have determined the numerical values of probabilities 
$$P(k,1)=\lim_{L\rightarrow\infty} P(k,1;L)$$  
for $k=1,2,$ and $3$ by means of finite-size scaling.
The values are given in Table~\ref{prob-sk1} for the cases of free
boundary conditions and periodic boundary conditions in the vertical
direction (i.e. perpendicular to the spanning direction).

\begin{table}
\centering
\caption{Probability to have at least $k$ horizontally spanning clusters 
in square with free boundaries (FBC) and periodic boundaries (PBC). Bond
critical percolation.}
\begin{tabular}{|l|l|l|} \hline
$n$ & $P(k,1)$ - FBC & $P(k,1)$ - PBC \\ \hline\hline
1   &  0.50002(2)     & 0.6365(1) \\
2   &  0.00658(3)     & 0.0020(4) \\
3   &  0.00000148(21) & 0.00000014(5) \\ \hline
\end{tabular}
\label{prob-sk1}
\end{table}

Actually, we use fully symmetric lattices, which are self dual even at
finite sizes. Therefore, relations (\ref{pi12}) holds for our finite
lattices (for details, see \cite{SK1}) and $P(1,1;L)\equiv 1$. In fact,
the data in Table~\ref{pap-1} support our observation. Finite size
corrections for the probabilities of at least 2 incipient spanning
clusters appear to be of the form $P(2,1;L)=P(2,1)+c/L^2$, with some
constant $c$, i.e. proportional to the inverse square lattice size $L$. We
found that the same finite size dependence holds also for $P(3,1;L)$
\cite{SK1}.

\begin{table}[htbp]
\centering
\caption{Probabilities $P(k,1;L)$ of at least $k$ incipient spanning clusters
on critical bond square lattices with size $(L,L)$ and free boundaries.
Note, that the values of $P(2,1;L)$ are multiplyed by a factor  $10^3$ and
those of $P(3,1;L)$ by a factor $10^6$. For each lattice size $L$ the
first row is the probability of horizontal spanning and the second one is
the probability of vertical spanning.}
\vspace*{13pt}
\begin{tabular}{r|r|r|r} \hline
$L$   & $P(1,1;L)$  & $P(2,1;L)\cdot 10^3$ & $P(3,1;L)\cdot 10^6$ 
\\ \hline \hline
8   & 0.50005(5) & 7.657(8)     & 3.40(15) \\ \hline
    & 0.50003(4) & 7.660(8)     & 3.98(21) \\ \hline\hline
12  & 0.50002(5) & 7.084(9)     & 2.57(14) \\ \hline
    & 0.49995(5) & 7.070(8)     & 2.10(13) \\ \hline\hline
16  & 0.50003(7) & 6.855(9)     & 1.97(17) \\ \hline
    & 0.50002(6) & 6.843(8)     & 1.79(19) \\ \hline\hline
20  & 0.49990(6) & 6.742(8)     & 1.95(14) \\ \hline
    & 0.50008(5) & 6.745(9)     & 1.72913) \\ \hline\hline
30  & 0.49999(4) & 6.650(8)     & 1.52(14) \\ \hline
    & 0.49996(5) & 6.653(7)     & 1.52(12) \\ \hline\hline
32  & 0.49999(5) & 6.648(8)     & 1.73(12) \\ \hline
    & 0.50008(7) & 6.642(8)     & 1.56(11) \\ \hline\hline
64  & 0.49992(9) & 6.597(9)     & 1.33(13) \\ \hline
    & 0.49999(6) & 6.602(8)     & 1.51(14) \\ \hline\hline
\end{tabular}
\label{pap-1}
\end{table}

The values of probabilities in Table~\ref{prob-sk1} tell us, that at
criticality there is percolation in one direction with the probability
equal to 1/2.  Among such samples and in the case of free boundaries, in
about 6 samples in 1000 there are two different clusters percolated
horizontally. Finally, 1 sample in 1000000 (one million) contain at least
3 clusters spanning lattice in the given direction. The corresponding
probabilities for $k>1$ are less in the case of periodic boundary
conditions, that is for cylindrical geometry.

The data presented in the Table~\ref{prob-sk1} appear in good agreement
with the exact form of the probabilities $P(k,r)$ conjectured by Cardy
using methods of conformal field theory \cite{Cardy98}. He extend the
arguments of his early paper \cite{Cardy92} and determined the exact
behaviour in the limit of infinite lattices $L\rightarrow\infty$ for the
spanning probabilities $P(k,r;L)$ for large aspect ratios $r$ in the case
of free boundaries

\begin{equation}
\lim_{L\rightarrow\infty} \ln P_{FBC}(k,r;L)
\propto -\frac{2\pi}{3} k(k-\frac12)\; r,
\label{CPF}
\end{equation}
as $r\rightarrow\infty$ for any $k$.

Analogously, for periodic boundaries in the vertical direction (spanning
along a cylinder) he found

\begin{equation}
\lim_{L\rightarrow\infty} \ln P_{PBC}(k,r;L)
\propto -\frac{2\pi}{3} (k^2-\frac14)\; r,
\label{CPP}   
\end{equation}
for $k\ge2$ and $r\rightarrow\infty$, and with the different exponent 
for $k=1$
\begin{equation}
\lim_{L\rightarrow\infty} \ln P_{PBC}(1,r;L)
\propto -\frac{5\pi}{24}\; r,
\label{CPP1}
\end{equation}
in the limit of large aspect ratio $r\rightarrow\infty$.

From the results (\ref{CPF}-\ref{CPP}) one could calculate ratio

$$
\frac{\ln P_{FBC}(k,1)}{\ln P_{PBC}(k,1)}=\frac{k(k-\frac12) }
{k^2-\frac14 },
$$
which is $\frac45=0.8$ for $k=2$ and $\frac67\approx.857$ for $k=3$.
Data from the Table~\ref{prob-sk1} give us the values $0.808(10)$ and 
$0.857(20)$, respectively. This close agreement with the asymptotic
form for $r\rightarrow\infty$ may be explained by the observation 
\cite{Cardy98} that the corrections to the leading behaviour are of order
$\exp(-2\pi r)$ (compare with expansion (\ref{z14ab})).

The detailed numerical check of Cardy's results (\ref{CPF}-\ref{CPP1})
is given in the remaining sections.

\section{Computational method.}

Our simulations can be divided into three major parts: 
\begin{description} 

\item[Sample choosing.] There are two ways to generate a sample. The first
one, is with the number of closed bonds (or sites) fixed, i.e. we choose
the sample from a {\em canonical ensemble}. This method is
especially convenient for the bond square percolation, when fully
self-dual lattices exist even for the finite system size \cite{SK1}. This
eliminates not only boundary effects, but also `fluctuations' of the
concentration.

The second way, is with a fixed probability for a given site (bond) to be
closed, i.e. sampling from a {\it grand canonical ensemble}. This method
is more convenient for the Ziff hull method, for the transfer-matrix
method and for the Hoshen-Kopelman algorithm we use in this work.

It is known that both methods lead to the same results in the
`thermodynamic' limit of infinite lattices \cite{Gr-Leb}. It should be
noted, that the computed averages coincide for both samples, whereas
corresponding dispersions are quite different, as was found recently
by Vasilyev and author \cite{VS}.

To generate samples, we generate random numbers as result of XOR
operations (eXclusive OR) applied to the output of two shift registers
with large length. Usually, we use the pairs of legs $(p,q)$ with
$(9689,471)$ and $(4423,1393)$, though some other combinations from
\cite{Heringa92} could be used as well. This give us an enormously large
period of random numbers and a lack of acting correlations. Details could
be found in papers \cite{SBH,SB,Ziff-CPC,LP} and references therein.

\item[Decomposition to clusters.] We use the Hoshen-Kopelman algorithm
\cite{HK} in their original ``single scan version''. That is we don't hold
the whole lattice in the memory but only the boundary column we start
from, the current column and the next column. In this way, the algorithm
can be named as a ``dimensionality reduction'' Hoshen-Kopelman algorithm.
This give us the possibility to fit all data together with the instruction
code within the 4 MB cache memory of Alpha workstations. So, we avoid the
slowest feature of modern computers, i.e. the slowing down of data flow
from CPU registers to RAM and back, typical for general simulations. The
resulting speed up of simulations is a factor of about $4$.

\item[Check of spanning.] We check spanning with the step $\Delta r$ in
aspect ratio equal to $1/4$ (and even $\Delta r=1/8$ for the largest width
$L=256$). Namely, the sample generation goes from right to left together
with the Hoshen-Kopelman labeling. After the successful generation of
$L/{\Delta r} =L/4$ sequential columns, we compare the cluster labels on
the start column with the cluster labels of the current column. The
resulting information is the indicator functions of the aspect ratio
$I_i(k,r)$, where $k$ denotes the number of independent spanning clusters
and $i$ is the sample number.  The value of this function at a fixed value
of variable $r$ is equal to 1 (there is spanning of $k$ clusters) or 0
(there is no spanning of $k$ clusters). This information is added to an
output file. After $10^6$ scans of the strip, the file is written onto the
disk and a new file is opened. The process was repeated $100$ times. So,
each of $100$ files contain the probabilities
$\pi_h(k,r;L)=10^{-6}\sum_{i=1}^{10^6} I_i(k,r)$ for the fixed value of
strip width $L$. Error bars are computed over 100 such values of
probabilities. It should be noted, that the probability $\pi_h(k,r;L)$ is
the probability of exactly $k$ clusters, and we have to compute $P(k,r;L)=
\sum_{j=k}^\infty \pi_h(j,r;L)$ (i.e. sum from $k$ to $\infty$) to get
spanning probabilities \cite{SK1} as defined in
Eqs.~(\ref{exact1},\ref{CPF}-\ref{CPP1}).

\end{description}
This ends the `measurement' of probabilities.

The detailed discussion of the computational methods will be published
elsewhere \cite{SK-fs}.

The remaining part of the lecture is devoted to the analysis of the
resulting data $P(k,r;L)$ \cite{SK1,SK-fs}.

\section{Finite size corrections of spanning probabilities. Free
boundaries.}
% Free boundaries and site percolation.

In this section we present an analysis of the probability $P(k,r;L)$ that
at least $k$ clusters span an open strip of width $L$ at least to a
distance $r$. Free boundary conditions are used throughout this section.  
An analysis of the probabilities with cylindrical geometry will be given
in the next section.

We generate $10^8$ different strips of width $L_v=L= $8, 12, 16, 20, 24,
28, 32, 48, 64, 128, 256 sites growing up to the length of $L_h=rL=32000$
sites. For the site percolation at $p_c=0.592746$ considered here about
$10^{15}$ random numbers were generated using `XOR' combination of the
output of two shift registers with the large lags. For some cases we
repeat simulations with the linear congruential {\tt rand48} generator
(compare with \cite{Lang}). Comparison did not demonstrate any evidence of
systematic errors.

As results of the simulations, we have data for $P(k,r;L)$. Our next task
is to determine the ISC exponents. For that, we have to take
$\lim_{L\rightarrow\infty}$ and determine the slope of $\ln P(k,r)$ as a
function of the aspect ratio $r$. This is the first method. Another one,
is to determine the slope of $\ln P(k,r;L)$ and then take the limit of
large strip widths $L\rightarrow\infty$.

Theoretically, both methods should give the same result. In practice the
two methods gave slightly different values which should be comparable
within error bars. Coincidence of both results could be a good test on the
data accuracy. Here we will demonstrate how both methods works for the
spanning probability $P(1,r;L)$ for which we have exact result
(\ref{exact1}) and then we will apply both methods to the probabilities of
multiple ISCs, for which only results for very large aspect ratios $r$ are
known (\ref{CPF}-\ref{CPP1}).

\subsection{First method.}

\begin{figure*}
\centering
\includegraphics[width=3in,height=3in]{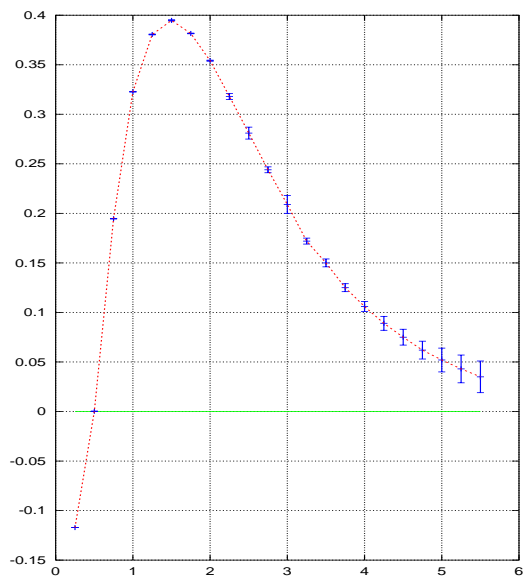}
\caption{Coefficient $b_0(r)$ in the fit to the spanning
probability in 2D site percolation as the function of the aspect
ratio $r$. See, Eqs.~\ref{lev-fs}.}  
\label{b0r}  
\end{figure*}

First, we have to develop finite size corrections of probabilities. It is
known from numerics, that the spanning probability for the site
percolation on square lattices with aspect ratio $r=1$ behaves at the
critical point as \cite{Ziff92}

\begin{equation}
P(1,1;L)\approx P(1,1)+\frac{b_0}{L+L_0}
\label{ziff-fs}
\end{equation}
with some constants $b_0=0.319$ and $L_0=1.6$.

Our data are in agreement with this result. We found that Ziff's proposal
could be generalized \cite{SK-fs} for any aspect ratio $r$
\begin{equation}
P(1,r;L)\approx P(1,r)+\frac{b_0(r)}{L+L_0(r)}
\label{lev-fs}
\end{equation}
where $b_0(r)$ and $L_0(r)$ are now some functions of the aspect ratio
$r$. $L_0(r)$ is a monotonic function vanishing very rapidly and can be
set to zero for the aspect ratio $r\gg 2.5$. Function $b_0(r)$ is shown on
Fig~\ref{b0r}. It is interesting that the spanning probability for the
site percolation does not depend on the lattice size for the aspect ratio
close to $1/2$. We could suggest that this aspect ratio $r=1/2$ be used
for the computation of percolation probability $p_c$ with a higher
accuracy than the usual simulations at $r=1$.

Our resulting function $P(1,r)$ coincides with the exact one
(\ref{exact1}) within error bars. The linear fit to the logarithm of the
resulting function $\ln P(1,r)$ in the interval $1.75\le r \le 5.5$ gives
the slop $\epsilon_1=-1.0476(3)$ which is very close to Cardy's
theoretical prediction $-\pi/3=-1.047197$.  To get an idea of the
accuracy, the linear fit to Ziff's approximation in the same interval was
done and yielded the slope $-1.0473(1)$. So, our first method produces
quite accurate data.

\subsection{Second method.}

Let us now compute first the slope $\epsilon_1(L)$ of the curves $\ln
P(1,r;L)$ with respect to the aspect ratio $r$ for each strip width $L$,
i.e. $\epsilon_1(L)$ and then take the limit

$$\lim_{L\rightarrow\infty} \epsilon_1(L)\rightarrow\epsilon_1.$$
 
Values of $\epsilon_1(L)$ are given in the Table~\ref{tsl-175}. 
A fit to the data in the table give 
\begin{equation}
\epsilon_1(L)=\epsilon_1+\frac{A}{L+L_0}=
-1.04703(2)+\frac{1.3416(1)}{L+1.7558(1)},
\end{equation} 
with the value of $\epsilon_1$ equal to the exact one in 4 digits.

It is interesting that the fit of values $\epsilon_1(L)$ with some
exponent $\theta$, i.e. with $\epsilon_1(L)=\epsilon_1+A/L^\theta$ yield
the result $\epsilon_1(L)=-1.05093(1)+0.8365(2)/L^{0.8522(1)}$ with too
large value of $\epsilon_1$ and the exponent $\theta=0.8522$ close to the
irrelevant exponent, introduced by Stauffer (see, f.e. \cite{Stauf-Ar})  
and included in the corrections to the scaling in the detailed analysis of
invariance in two dimensional percolation \cite{HoviAr}. Our data does not
support this exponent but rather the expansion in powers of $1/L$ which
could be effectively written as the term $A/(L+L_0)$. Such a term was used
first in a paper by Levinstein, et al. (see, f.e.
\cite{Efros-s})\footnote{We are glad to acknowledge discussion of this
point with E.I.~Rashba and A.L.~Efros, who stressed that they was
introduced this term ``to make the line as straight as possible``.} in
their first-time computation of the correlation length exponent and
successfully used later in the context of spanning probabilities by Ziff
\cite{Ziff92} in his computation of the percolation threshold value
$p_c=0.592746$ for the site percolation, which is the most precise value
(see also \cite{Ziff-fs}). We found this form to be very helpful in the
data analysis.

\begin{table}
\centering
\caption{Slope $\epsilon_1(L)$ of the $\ln P(1,r;L)$ as a function of the
aspect ratio
$r=L_h/L_v$ for several strip widths $L_v=L$. Slopes are computed in the
range $1.75\le r \le 5.5$.}
\begin{tabular}{rrl}
$L$  & $\epsilon_1(L)$ & $\sigma$ \\ 
8  &  -0.909926  &   .0001 \\
12 &  -0.949355  &   .00003 \\
16 &  -0.971482  &   .0001 \\
20 &  -0.985734  &   .0001 \\
24 &  -0.995344  &   .00005 \\
28 &  -1.001989  &   .00008 \\
32 &  -1.007151  &   .0001  \\
48 &  -1.019926  &   .00004  \\
64 &  -1.026430  &   .00007  \\
128&  -1.036904  &   .00007  \\ \hline
$\infty$ & -1.047032 & .00002 
\end{tabular}
\label{tsl-175}
\end{table}

\begin{figure*}
\centering
\includegraphics[width=3in,height=3in]{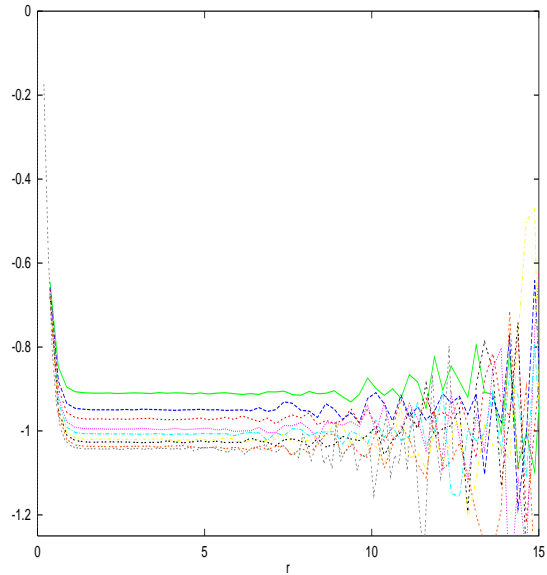}
\caption{Derivative of the logarithm of the spanning probability
$P(1,r;L)$ as a function of the aspect ratio $r$ for the strip width
$L=8,12,16,24,32,48,64,128,256$ from top to bottom. Site percolation on
a square open strip.}
\label{der-1}
\end{figure*}

The second method could be modified in the following way.
Taking the derivative of $\ln P(k,r)$ with respect
to aspect ratio $r$ we get
\begin{equation}
\frac{d}{dr}\ln P(k,r)\propto -\frac{2\pi}{3} k(k-\frac12)=\epsilon_k.
\label{lnP}
\end{equation}

\begin{figure*}
\centering
\includegraphics[width=3in,height=3in]{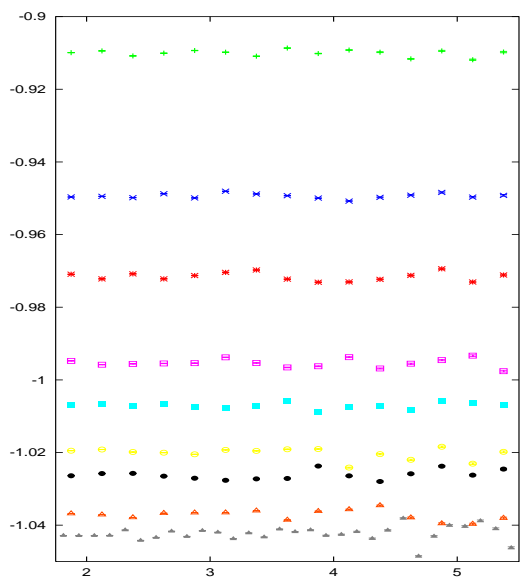}
\caption{The same as Fig.\ref{der-1} for the aspect ratio in the region 
$1.75\le r\le 5.5$. Derivative of the logarithm of the spanning 
probability $P(1,r;L)$ for strip widths $L=8,12,16,24,32,48,64,128,256$
from top to bottom.}
\label{all-25}
\end{figure*}

This give us the idea the to compute $\epsilon_k(r,L)$ taking the
numerical derivative of our data $\ln P(k,r;L)$. The result is shown on
Fig.~\ref{der-1} for site percolation in a strip with several widths $L$.
The slope of $\ln P(1,r;L)$ is constant in some region of $r$. Deviation
below the left-hand value $r_{min}\approx 1.5$ is due to the corrections
$\exp(-2\pi r)$ to the leading behavior as seen from Ziff's approximation
(\ref{z14ab}) and becomes negligibly small starting from $r=1.75$. Large
fluctuations for aspect ratios larger than $r_{max}\approx 5.5$ is due to
the small values of probabilities $P(1,r;L)$ for large $r$ and, therefore,
error bars become of the size of the fluctuations (not shown on the
Fig.~\ref{der-1} for the reason of picture clarity). The Fig.~\ref{all-25}
shows a magnified picture with the same derivative for the values of the
aspect ratio $1.75\le r\le 5.5$ together with the error bars.

We now take the simple average of $\epsilon_1(r,L)$ over $r$ in the
mentioned region and get values of $\epsilon_1(L)$. In the same manner as
before, we could now take $\epsilon_1=\lim_{L\rightarrow\infty}
\epsilon_1(L) \approx 1.0473(1)$ which is consistent with the values
obtained by other methods and coincides well with the exact value. 

It should be noted that the accuracy of this method is not as good as that
of the previous ones because we take simple numerical derivatives, which
magnify all fluctuations. Nevertheless, the method of derivatives give us
clear idea on the interval $(r_{min},r_{max})$ of aspect ratio over which
we can safely compute exponents. So, it is helpful as a supplement to the
two methods described above.

\subsection{Two and three ISC's in an open strip.}

The same analysis of the data for probability of at least two ISC leads to
the exponent $\epsilon_2=-6.294(6)$ which is within two sigma from the
exact value $-2\pi=-6.2832$. The corresponding exponent for the three
ISC's are also computed $\epsilon_3=-15.51(1)$. This value is well
coincides with the exact value $-28/3\pi\approx -15.708$. We got
relatively large deviation because the value of such probability decays
very fast with the aspect ratio $r$ and practically we take averages only
over the rather small interval of aspect ratio $0.5\le r \le 1.25$. The
good agreement could be explained by the small corrections of order of
$\exp(-2\pi r)$ to the leading behaviour \cite{Cardy98}.

\section{Periodic boundary conditions.}

Periodic boundary conditions in vertical directions lead to different
values of spanning probabilities. The geometry is equivalent to an open
ended cylinder with the circle length $L_v=L$ and ``horizontal'' length
$L_h=rL$. Clusters can span from the one cylinder end to another one also
rotating around the cylinder. This leads to larger values of the one
cluster spanning probability than in the case of open strip described in
above. Actually, we obtain a spanning probability equal to $0.6365(1)$
which is larger than $0.5$ in the case of the open strip and in a good
comparison with the result of the paper \cite{HoviAr} and more accurate.  
So, the spanning probability depends not only on the aspect ratio but on
the boundary conditions as well \cite{Lang,Lang-Bull,HoviAr}. As a
consequence, the values of multiple ISC probabilities should be smaller
than the corresponding ones for the open strip.

\begin{figure*}
\centering
\includegraphics[width=3in,height=3in]{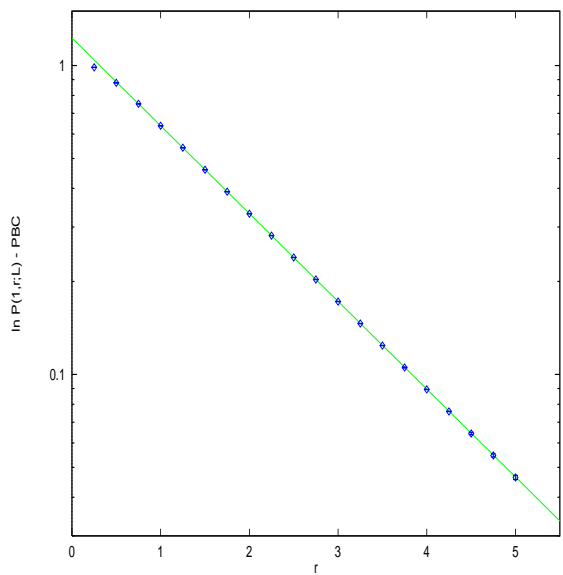}
\caption{Spanning probability $P(1,r)$ along cylinder as function of
aspect ratio $r$ denoted by open rhombs with error bars. Solid line is the
best fit to probability in the interval $1.75\le r \le 5.5$.}
\label{uga1}
\end{figure*}

We applied both methods described in the previous section to the data
analysis of the spanning probability $P(1,r;L)$ along an open ended
cylinder.  

Using the first method we compute the slope $\epsilon_1$ of the logarithm
of the spanning probability and get $\epsilon_1=-0.6547(2)$.
Fig.~\ref{uga1} shows the probability together with the best fit.

The second method based on the slope $\epsilon_1(L)$ calculation for each
given lattice size $L=8,12,16,20,24,28,32,36,40,48,64$ and then taking the
limit of infinite lattice, finally leads to the value
$\epsilon_1=-0.65448(5)$. Both values are very close to the exact value
$-5\pi/24=-0.654498$. The second method produce much better accuracy.

Finite size corrections to the limiting values of the slope are similar to
those for free boundaries. The only difference is in the finite size
corrections of $\epsilon_1(L)$. The leading term, proportional to the
inverse width vanishes, and the second power plays the main role, namely
$\epsilon_1(L)=\epsilon_1-0.309(6)/L^2$. (We recall, that in the case of
open boundaries the finite size corrections were
$\epsilon_1(L)=\epsilon_1+c/(L+L_0)$.)

The $1/L^2$ finite size corrections also hold for $\epsilon_2(L)$ and
$\epsilon_3(L)$, whose values are $\epsilon_2(L)=-7.852(1)$ and
$\epsilon_3(L)=-18.11(15)$, also quite close to the exact values
$-5\pi/2=-7.8540$ and $-35\pi/6=-18.33$.

\section{Incipient spanning clusters in simple cubic lattice}

Aizenman, in his seminal paper \cite{Aiz1}, also proposed the ISC
exponents in dimension $d$ from lower $d_l=2$ to upper $d_u=6$ critical
dimensions

\begin{equation}
P_h(k,r)\propto \exp(-\alpha k^\frac{d}{d-1}) =\exp(-\epsilon_k r).
\label{exp-d}
\end{equation}

The preliminary analysis of the data obtained for the probabilities of
clusters spanning between two opposite planes of the open cube $(rL,L,L)$
at the critical site percolation threshold $p_c=0.3116$
\cite{Gras-3d,JanSt} are presented in the Table~\ref{3d}.  We compute
values of $\epsilon_k$ using the second method described above for
system sizes $L=8,16,32,48,64$ and aspect ratio up to $r=7.25$, so some
samples contain 1900544 sites. The averages were taken over $10^7$
such samples.

 A possible fit to the data in Table~\ref{3d} is $\alpha_k\propto -\pi
(k^{1.5}-0.22 k+0.11)$, which supports Aizenman proposal. We cannot
exclude, nevertheless, the power in exponent being less than
$d/(d-1)=1.5$.

In their preprint \cite{SenBha}, Sen and Bhattacharjee proposed on the
basis of their simulations that in the directed 3D percolation the ISC
exponent depends with the same power $2$ on the number of clusters $k$.
Here we compute directly exponents as the slope of the logarithm of the
corresponding probabilities, whereas in the Sen and Bhattacharjee preprint
only values of probabilities for one fixed aspect ratio $r=1$ were
determined. We are not sure their analysis could distinguish the powers
$1.5$ and $2$. We could only say that analysing our very precise data, we
believe the value of power to be not larger than $1.5$.

More detailed simulations and analysis are in progress.

\begin{table}
\center
\caption{Exponents $\epsilon_k$ depending on the number $k$ of ISC
clusters in simple cubic site critical percolation.}
\begin{tabular}{|l|l|l|} \hline
$k$ & $\epsilon_k$ & $\sigma$ 
\\ \hline\hline
1 & -2.7641 & 0.0004 \\
2 & -7.935  & 0.001 \\
3 & -14.68   & 0.01 \\
4 & -21.7   & 0.2 \\ \hline
\end{tabular}
\label{3d}
\end{table}

\section{Discussion}

We compute the values of the new exponents connected with the
probabilities of a number of incipient spanning clusters in two
dimensional percolation at criticality.

The results of comparison of the computed values with the exact ones are
summarized in Table~\ref{tab-res}. Our data strongly support the exact
results (\ref{CPF}-\ref{CPP1}) for the number of incipient spanning
clusters in two dimensional percolation developed by Cardy in
\cite{Cardy92,Cardy98} and, therefore, the validity of conformal
invariance for multiple spanning probabilities.

\begin{table}
\centering
\caption{Comparison of computed data with the exact results. Site
percolation in two dimensional square lattice with free boundaries (FBC)
and periodic boundaries (PBC).}
\begin{tabular}{|l|ll|ll|}\hline
 & \multicolumn{2}{|c|}{Free boundaries} &
\multicolumn{2}{c|}{Periodic boundaries} \\ \cline{2-5}
$k$ & \multicolumn{1}{c|}{$\alpha_k$ Exact }&
\multicolumn{1}{c|}{$\alpha_k$
Comp.} 
& \multicolumn{1}{c|}{$\alpha_k$ Exact }&
\multicolumn{1}{c|}{$\alpha_k$
Comp.} \\ \hline
1 & $-\frac{\pi}{3}=-1.047197$ & -1.0473(2) & 
  $-\frac{5\pi}{24}=-0.654498$ & -0.65448(5)  \\
2 &  $-2\pi=-6.28318$ & -6.294(6) & 
  $-\frac{5\pi}{2}=-7.8540$ & -7.852(1)  \\
3 & $-5\pi=-15.70796...$ & -15.51(1) &
   $-\frac{35\pi}{6}=-18.33$ & -18.11(15)   
\\ \hline
\end{tabular}
\label{tab-res}
\end{table}

Finite size corrections of the data computed with a high accuracy
demonstrate that at criticality, corrections to all quantities
analyzed depends on the inverse system size or on the inverse square size.
We did not find any evidence for the irrelevant exponent $\theta_1=0.85$
as discussed in \cite{HoviAr}.

We stress the possibility of self duality for {\em finite} square lattices
in the case of bond percolation.

We generalize finite size corrections developed by Ziff for the site
percolation in a square geometry to any aspect ratio. We found that finite
size scaling of the probability of cluster spanning along a strip with
open boundaries is described by a nonmonotonic function $b_0(r)$ which
vanishes at the aspect ratio $r=1/2$, or nearby.

Our preliminary data support Aizenman's proposal for the ISC exponent in
three dimensional critical percolation. More data and analysis are
necessary for definite check.

\section{Acknowledgments}

The author would like to thank D.~Stauffer, A.~Aharony, R.~Ziff, J.~Cardy,
P.~Butera and J.-P.~Hovi for fruitful discussions. Special thanks to
S.~Kosyakov together with whom most of the results were computed. The
author is thankful for the invitation to D.P.~Landau and for kind
hospitality to the Center of Simulational Physics, University of Georgia,
where this paper was finished.  This work was supported in part by grants
from RFBR, INTAS, Centro Volta - Landau Network and NWO.


\begin{thebibliography}{999}

\bibitem{Bunde}  A. Bunde, S. Havlin, Eds.,
{\it Fractals and Disordered Systems}, (Springer, Berlin, 1996)

\bibitem{FK} C.M. Fortuin, P.W. Kasteleyn, Physica {\bf 57}
(1972) 536

\bibitem{Stauf-Ar} D. Stauffer, A. Aharony.
{\it Introduction to Percolation Theory}, 2nd ed.
(Taylor \& Francis, London, 1995)

\bibitem{Lang} R. Langlands, C. Pichet, P. Pouliot, Y. Saint-Aubin,
J. Stat. Phys. {\bf 67}  (1992) 553

\bibitem{Ziff92} R.M. Ziff, Phys. Rev. Lett. {\bf 69} (1992) 2670

\bibitem{Aiz1} M. Aizenman, Nucl. Phys. B [FS] {\bf 485} (1997) 551

\bibitem{Grim} G. Grimmett, {\it Percolation}, (Springer, New York, 1989)

\bibitem{Hu} C.-K. Hu and C.-Y. Lin, Phys. Rev. Lett., {\bf 77} (1996) 8

\bibitem{SK1} L.N. Shchur and S.S. Kosyakov,
             Int. J. Mod. Phys. C 8 (1997) 473,
	     Nucl. Phys. B (Proc. Suppl.) {\bf 63A-C} (1998) 664

\bibitem{Sen} P. Sen, Int. J. Mod. Phys. C  {\bf 7} (1996), Int. J. Mod.
Phys. C {\bf 8} (1997) 229

\bibitem{Cardy98} J. Cardy, J. Phys. A {\bf 31} (1998) L105 

\bibitem{Topos} S.S. Kosyakov, S.A. Krashakov, L.N.
Shchur,  Proc. Int. Conf. PDPTA'97, Las Vegas, Nevada, USA, Vol.2,
1170-1173 (Ed. H. R. Arabnia, 1997)

\bibitem{Essam-80} J. W. Essam, Rep. Prog. Phys. {\bf 43} (1980) 833

\bibitem{Kesten} H. Kesten, {\it Percolation theory for mathematicians}, 
(Birkh\"auser, Boston, 1982) 

\bibitem{AizB} M. Aizenman and D.J. Barsky, Comm. Math. Phys. {\bf 108}
(1987) 489

\bibitem{RKS} P.J. Reynolds, H.E. Stanley, and W. Klein, 
J.Phys. A {\bf 11} (1978) L199, Phys. Rev. B {\bf 21} (1980) 1223

\bibitem{Cardy92} J.L. Cardy,
J. Phys. A {\bf 25} (1992) L201

\bibitem{Heringa92}
J.R. Heringa, H.W.J. Bl\"ote, A. Compagner,
Int. J. Mod. Phys. C  {\bf 3} (1992) 561

\bibitem{Lang-Bull} R. Langlands, P. Pouliot, Y. Saint-Aubin,
Bulletin (New Series) of the American Mathematical
 Society, {\bf 30} (1994) 1

\bibitem{mini-review} D. Stauffer,
Physica A {\bf 242}, (1997) 1

\bibitem{Ziff-C} R.M. Ziff, J. Phys. A {\bf 28} (1995) 1249
, J. Phys. A {\bf 29} (1995) 6479

\bibitem{HoviAr}  J.-P. Hovi, A. Aharony,
Phys. Rev. E {\bf 53} (1996) 235

\bibitem{Aiz-19} M. Aizenman,
in proceedings of STATPHYS 19, editor Hao Bailin (World Scientific,
Singapore, 1996)

\bibitem{Gr-Leb} R.B. Griffiths and J.L. Lebowitz,
J. Math. Phys. {\bf 9} (1968) 1284

\bibitem{VS} O.A. Vasilyev and L.N. Shchur, unpublished

\bibitem{SBH}
   L.N. Shchur, H.W.J. Bl\"ote and J.R. Heringa,
   Physica A {\bf 241} (1997) 579

\bibitem{SB}
   L.N. Shchur and H. Bl\"ote,
   Phys. Rev. E  {\bf 55} (1997) R4905

\bibitem{Ziff-CPC} R.M. Ziff,
                    Computers in Physics {\bf 12} (1998) 385

\bibitem{LP} L.N. Shchur and P. Butera, Int. J.Mod. Phys C {\bf 9} (1998)
607

\bibitem{HK} J. Hoshen, R. Kopelman,
Phys. Rev. B {\bf 14}  (1976) 3438

\bibitem{SK-fs} L.N. Shchur and S.S. Kosyakov, unpublished

\bibitem{Efros-s}  B.I. Shklovskii, A.L. Efros,
{\it Electronic Properties of Doped Semiconductors. 5. Percolation
Theory} (Springer-Verlag, Berlin, 1984) pp. 95-136

\bibitem{Ziff-fs} R.M. Ziff, Phys. Rev. E {\bf 54} (1996) 2547

\bibitem{Gras-3d} P. Grassberger, J. Phys. A {\bf 25} (1992) 5867

\bibitem{JanSt} N. Jan and D. Stauffer, Int. J. Mod. Phys. {\bf C9}
(1998) 341

\bibitem{SenBha} P. Sen and S.M. Bhattacharjee, preprint cond-mat/9804282

\end{thebibliography}
\end{document}